\documentclass[11pt]{article}
\usepackage{authblk}
\usepackage{graphicx}
\usepackage{color, colortbl}
\usepackage[caption=false,font=footnotesize]{subfig}
\usepackage{amsmath, amssymb}
\usepackage{enumitem}
\usepackage{multirow}

\usepackage{float}

\usepackage{lineno}

\begin{document}

\title{Graph Distances and Clustering}

\author[1]{Pierre Miasnikof \thanks{corresponding author: p.miasnikof@mail.utoronto.ca}}
\author[2]{Alexander Y. Shestopaloff}
\author[3]{Leonidas Pitsoulis}
\author[1]{Yuri Lawryshyn}

\affil[1]{University of Toronto, Toronto, ON, Canada}
\affil[2]{The Alan Turing Institute, London, United Kingdom}
\affil[3]{Aristotle University of Thessaloniki, Thessaloniki, Greece}

\date{}

\maketitle

\begin{abstract}
With a view on graph clustering, we present a definition of vertex-to-vertex distance which is based on shared connectivity. We argue that vertices sharing more connections are closer to each other than vertices sharing fewer connections. Our thesis is centered on the widely accepted notion that  strong clusters are formed by high levels of induced subgraph density, where subgraphs represent clusters. We argue these clusters are formed by grouping vertices deemed to be similar in their connectivity. At the cluster level (induced subgraph level), our thesis translates into low mean intra-cluster distances. Our definition differs from the usual shortest-path geodesic distance. In this article, we compare three distance measures from the literature. Our benchmark is the accuracy of each measure's reflection of intra-cluster density, when aggregated (averaged) at the cluster level. We conduct our tests on synthetic graphs generated using the planted partition model, where clusters and intra-cluster density are known in advance. We examine correlations between mean intra-cluster distances and intra-cluster densities. Our numerical experiments show that Jaccard and Otsuka-Ochiai offer  very accurate measures of density, when averaged over vertex pairs within clusters.
\end{abstract}

\section{Introduction}
When clustering graphs, we seek to group nodes into clusters of nodes that are similar to each other. We posit that similarity is reflected in the number of shared connections. On the basis of this shared connectivity, we establish node-to-node distances. These distances are inversely related to similarity, to shared connectivity.

Although a formal definition of vertex clusters or node communities remains a topic of debate (e.g., \cite{guideFortunato16}), virtually all authors agree a cluster (or community) is a subset of vertices that exhibit a high-level of interconnection between themselves  and a low-level of connection to vertices in the rest of the graph \cite{FortunatoLong2010,YangLesko2012, modWAW2016,EuroComb2017} (we quote these authors, but their definition is very common across the literature). Consequently, strongly inter-connected sets of vertices also form dense induced subgraphs. In line with this virtually universal agreement, we compare the accuracy of various node-to-node distance measures in reflecting intra-cluster density. The choice of intra-cluster density as a benchmark is consistent with this widely accepted definition of clusters.

\section{Distance, Intra-Cluster Density and Graph Clustering (Network Community Detection)}
As mentioned previously, clusters are defined as subsets of vertices that are considered somehow similar. This similarity is captured by the number of shared connections and translated into distance. In our model, vertices sharing a greater number of connections are closer to each other than to vertices with which they share fewer connections. It is important to note here that, in our definition, distance measures similarity, not geodesic (shortest path) distance. For example, two vertices that share an edge but no other connection have a geodesic distance of one, but they are arguably dissimilar.

At the cluster level, this distance takes the form of subsets of densely connected vertices. The link between clustering and density has been discussed in depth, recently  \cite{PMEtAlWAW18,PMEtAl2019,PMEtAl8ICN,PMEtAl2020}. In this article, our ultimate goal is to transform a graph's adjacency matrix into a $ \vert V \vert \times \vert V \vert$ similarity or distance matrix $D=\left[ d_{ij} \right]$, where the distance between each pair of vertices is given by the element $d_{ij}$. This transformation allows us to use the quadratic formulation of the clustering problem proposed by Fan and Pardalos \cite{FanPard2010CND,FanPard2010LQ}. Such a formulation can then be further modified into a QUBO formulation \cite{Glover2019}, which can be implemented on purpose-built hardware, like Fujitsu's Digital Annealer \cite{physInspired2019,applic2019}, for example. This purpose built architecture allows us to circumvent the NP-hardness of the clustering problem \cite{Fu86,Schaeffer2007,FortunatoLong2010,IsingForm2014,physInspired2019}.

\begin{figure}
	\centering
	\includegraphics[width = 0.5\textwidth]{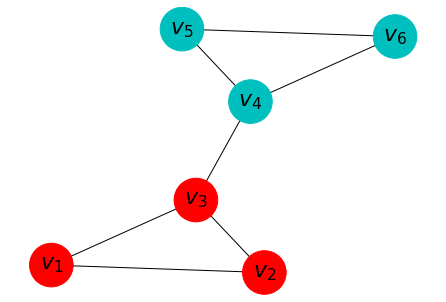} 
	\caption{Graph with Two Clusters}
	\label{example}
\end{figure} 
To illustrate our definition of distance, we examine the graph shown in Figure~\ref{example}. The graph in that figure is arguably composed of two clusters (triangles), the red cluster containing vertices $v_1,v_2,v_3$ and the cyan cluster with vertices $v_4,v_5,v_6$. We observe that each cluster forms a dense induced subgraph (clique). We also note that the geodesic distance separating vertices $v_1$ and $v_3$ is equal to the geodesic distance separating $v_3$ and $v_4$. Nevertheless, in the context of clustering, we argue that $v_3$ is closer, more similar, to $v_1$ than to $v_4$. The ultimate goal of this study is to identify a distance measure that accurately measures this similarity in connectivity.

\section{Distance Measurements Under Study}
We compare three different distance measurements from the literature and examine how faithfully they reflect connectivity patterns. We argue that mean node-to-node distance within a cluster should offer an accurate reflection of intra-cluster density. Intra-cluster density, defined as
\[
K_{\text{intra}}^{(k)} = \frac{ \vert E_{kk} \vert}{0.5 \times n_k \times (n_k - 1)} \, .
\]
In this definition, $\vert E_{kk} \vert$ is the cardinality of the set of edges connecting two vertices within the same cluster `$k$' and $n_k = \vert V_k \vert$ is the number of vertices in that same cluster.

We then examine the relationship between mean Jaccard \cite{JaccOrig}, Otsuka-Ochiai \cite{Ochiai57} and Burt's distances \cite{BurtDist76}, on one hand, and intra-cluster density \cite{PMEtAlWAW18,PMEtAl8ICN,PMEtAl2019,PMEtAl2020} within each cluster, on the other. Because these distances are pairwise measures, we compare their mean value for a given cluster to the cluster's internal density.

\subsection{Jaccard Distance}
The Jaccard distance separating two vertices `$i$' and `$j$' is defined as
\[
\zeta_{ij} = 1 - \frac{ \vert c_i \cap c_j \vert  }{ \vert c_i \cup c_j \vert} \in [0,1] \, .
\]
Here, $c_i \, (c_j)$ represents the set of all vertices with which vertex `$i \, (j)$' shares an edge. 

At the cluster level, we compute the mean distance separating all pairs of vertices within the cluster, which we denote as $\mathcal{J}$. For an arbitrary cluster `$k$' with $n_k$ vertices, we have
\[
\mathcal{J}_k = \frac{1}{0.5 \times n_k \times (n_k -1)} \sum_{i,j=i+1} \zeta_{ij} \, .
\]

\subsection{Otsuka-Ochiai Distance}
The Otsuka-Ochiai (OtOc) distance separating two vertices `$i$' and `$j$' is defined as
\[
o_{ij} = 1 -  \frac{\vert c_i \cap c_j \vert}{ \sqrt{\vert c_i \vert \times \vert c_j \vert} } \in [0,1] \, .
\]
Here again, to obtain a cluster level measure of similarity, we take the mean over each pair of nodes within a cluster. We denote this mean as $\mathcal{O}$. Again, for an arbitrary cluster `$k$' with $n_k$ vertices, we have
\[
\mathcal{O}_k = \frac{1}{0.5 \times n_k \times (n_k -1)} \sum_{i,j=i+1} o_{ij} \, .
\]

\subsection{Burt's Distance}
Burt's distance between two vertices `$i$' and `$j$' is computed as
\[
b_{ij} = \sqrt{ \sum_{k \ne i,j} \left( A_{ik} - A_{jk} \right)^2 } \, .
\]
At the cluster level, we denote the mean Burt distance as $\mathcal{B}$. For an arbitrary cluster `$k$' with $n_k$ vertices, it is computed as
\[
\mathcal{B}_k = \frac{1}{0.5 \times n_k \times (n_k -1)} \sum_{i,j=i+1} b_{ij} \, .
\]

\section{Numerical Comparisons}
To compare the distance measures, we generate synthetic graphs with known cluster membership, using the planted partition model. Then, for each of our test graphs, we compute our three vertex-to-vertex distances. We then compute mean distances between nodes in each cluster and intra-cluster density.

To assess the accuracy of each measure as a reflection of intra-cluster density, we examine the (Pearson) correlation between each distance measure and intra-cluster density. We examine correlations for each graph and for the set of all graphs. We also record correlations between each mean distance measure and the probability of inter-cluster connection used to generate these graphs. 

\subsection{Test Data: Synthetic Graphs with Known Clusters}
We use the planted partition model to generate the 10 graphs described in Table~\ref{graphs}. All graphs consist of 50 clusters. We vary clusters sizes across graphs, but these sizes are kept constant within each graph. We also vary the edge probability within clusters and between vertices in different clusters, but keep them constant across clusters and cluster pairs, as per the planted partition model. These graphs were generated using the Python NetworkX library \cite{Networkx}. Intra- and inter-cluster  edge probabilities as well as cluster sizes are shown in Table~\ref{graphs}.

\begin{table}[]
	\centering
	\caption{Synthetic Graphs and their Characteristics}
	\label{graphs}
	\begin{tabular}{rrrrr}
		\hline
		Graph & Intra Pr & Inter Pr & $n_k$ & $\vert V \vert$   \\
		\hline
		G1    & 1             & 0             & 45   & 2,250 \\
		G2    & 0.9           & 0.1           & 37   & 1,850 \\
		G3    & 0.9           & 0.15          & 42   & 2,100 \\
		G4    & 0.9           & 0.2           & 50   & 2,500 \\
		G5    & 0.8           & 0.1           & 53   & 2,650 \\
		G6    & 0.8           & 0.15          & 38   & 1,900 \\
		G7    & 0.8           & 0.2           & 44   & 2,200 \\
		G8    & 0.7           & 0.1           & 39   & 1,950 \\
		G9    & 0.7           & 0.15          & 46   & 2,300 \\
		G10   & 0.7           & 0.2           & 53   & 2,650 \\
		\hline
	\end{tabular}
\end{table}

\subsection{Empirical Results}
As expected, in the case of graph G1 where the intra-cluster edge probability is one and inter-cluster edge probability is zero, the case of a graph composed of disconnected cliques, all three distances are constant across all clusters. In this case, all vertices within each cluster have exactly the same connectivity pattern (same neighbors). They are all separated by the same distances. Therefore, correlation to intra-cluster density is meaningless. These distances are recorded in Table~\ref{disconnected}.

\begin{table}[]
	\centering
	\caption{Mean Distances in Disconnected Cliques}
	\label{disconnected}
	\begin{tabular}{|rrr||rrr|}
		\hline
		Graph & Intra Pr & Inter Pr & $\mathcal{J}$ & $\mathcal{B}$ & $\mathcal{O}$  \\
		\hline
		G1 & 1 & 0 & 0.02 & 0 & 0.04 \\
		\hline
	\end{tabular}
\end{table}
While the case of graph G1 is predictable, it is important to note that Jaccard and OtOc distances are not zero, in the case of a complete (sub)graph with no self-loops. This difference is due to the numerator of these quantities. In the case of complete graphs with no self loops, a node `$v_i$' is connected to node `$v_j$' but not to itself. As a result we have the following inequality of the cardinalities: $\vert c_i \cap  c_j \vert <  \vert c_i \cup  c_j \vert$, for any pair of vertices in a complete (sub)graph.

For our other graphs (G2-G10), the exact number of intra- and inter-cluster edges are probabilistic. As a result, all distances and densities are random variables. This randomness allows a comparison of (Pearson) correlations between the distances and intra-(inter-)cluster densities. However, before performing these comparisons, we examine the relationship between distances and intra-(inter-)cluster densities, graphically.

\begin{figure}
	\centering
	\subfloat[Mean Intra-Cluster Jaccard Distances $(\mathcal{J})$]{ \includegraphics[width = 0.5\textwidth]{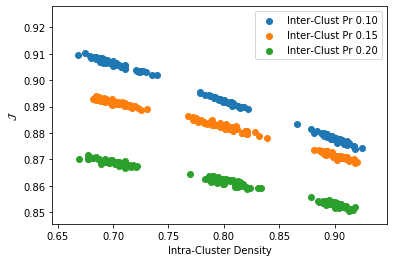}   \label{Jplot} } \\
	\subfloat[Mean Intra-Cluster OtOc Distances $(\mathcal{O})$]{ \includegraphics[width = 0.5\textwidth]{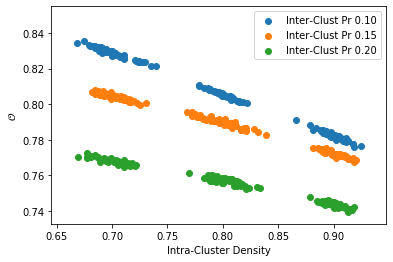} \label{Oplot}  } \\
	\subfloat[Mean Intra-Cluster Burt Distances $(\mathcal{B})$]{ \includegraphics[width = 0.5\textwidth]{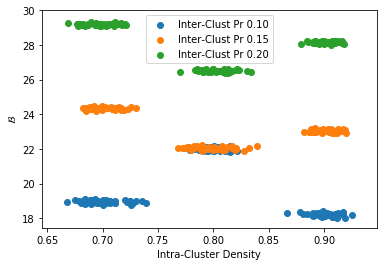} \label{Bplot}  }
	\caption{Mean Intra-Cluster Distances as a Function of Inter- and Intra-Cluster Densities}
	\label{plots}
\end{figure} 

Upon examining Figure~\ref{plots}, we immediately note the distances $\mathcal{J}$ and $\mathcal{O}$ have a linear negative relationship with intra-cluster density. This strongly linear relationship justifies the use of Pearson correlation coefficients. Meanwhile, Burt's distance $(\mathcal{B})$ seems only loosely related to intra-cluster density, at best.

It is also interesting to note that mean Jaccard and OtOc distances across the graph decrease with increases in inter-cluster edge probability. In contrast, Burt's distances decrease with inter-cluster edge probability. These trends can be observed for distances within clusters in Figure~\ref{plots}. In Table~\ref{alldists}, we also observe the same phenomenon for all distances across the graph, regardless of cluster membership.  

\begin{table}[]
	\centering
	\caption{Mean Distance, Cluster Membership Not Considered}
	\label{alldists}
	\begin{tabular}{|r||rrr|}
		\hline
		Inter-Clust Pr & $\mathcal{J}$    & $\mathcal{B}$     & $\mathcal{O}$    \\
		\hline 
		0              & 0.04 & 0     & 0.02 \\
		0.10           & 0.89 & 19.72 & 0.81 \\
		0.15           & 0.88 & 23.15 & 0.79 \\
		0.20           & 0.86 & 27.95 & 0.76 \\
		\hline
	\end{tabular}	
\end{table}

Intuitively, Burt's distance increases with the probability of inter-cluster connection. As this probability increases, nodes share a smaller proportion of their connections. Meanwhile, the trends observed in $\mathcal{J}$ and $\mathcal{O}$ are direct consequences of their mathematical definition:
\begin{eqnarray*}
	\zeta_{ij} &=& 1 - \frac{\vert  c_i \cap c_j \vert }{ \vert c_i \cup c_j \vert} \\
	o_{ij} &=& 1 -  \frac{\vert c_i \cap c_j \vert }{ \sqrt{\vert c_i \vert \times \vert c_j \vert} } \, .
\end{eqnarray*}
In both cases, the numerator is the number of shared connections. The denominator is proportional to all connections of either vertex `$i$' or `$j$' and increases at a much higher rate. For example, with all else equal, as the probability of inter-cluster connection increases from 0, the total number of connections (degree) of both vertices `$i$' and `$j$' increases sharply, at a mean rate of the order of $2 \times P_{\text{inter}} \times \left( \vert V \vert - n_k \right)$. However, the numerators, which correspond to the number of shared connections, increase at a much lower mean rate. They increase at a mean rate of $P_{\text{inter}}^2 \times \left( \vert V \vert - n_k \right)$.

Nevertheless, we note a clear linear inverse relation between intra-cluster density and both $\mathcal{J}$ and $\mathcal{O}$. This relationship is also observed in the correlations, shown in Table~\ref{corrs}.

\begin{table}[]
	\centering
	\caption{Correlation Coefficient Between Distance and Intra-Cluster Density} 
	\label{corrs}
	\begin{tabular}{|r|rrr|}
		\hline
		& \multicolumn{3}{c|}{$\rho$ to Intra-Clust Density} \\
		Inter Pr & $\mathcal{J}$  &  $\mathcal{B}$  & $\mathcal{O}$      \\
		\hline
		0			& NA & NA & NA \\
		0.10            & -0.999 & -0.182 &-0.999 \\
		0.15           & -0.997 & -0.551 &-0.997 \\
		0.20            & -0.993 & -0.400 &-0.994 \\
		\hline
		All $(\ne 0)$ &  -0.563 & -0.116 & -0.565 \\
		\hline
	\end{tabular}	
\end{table}

\section{Our Chosen Distance}
Both Jaccard and OtOc distances are very accurate reflections of intra-cluster density. When averaged over all vertices within clusters, they exhibit almost perfect inverse correlation to intra-cluster density. 

However, the Jaccard similarity and it's complement, the Jaccard distance, are used widely in a variety of different fields. Because of its widespread use and the availability of pre-built computational functions, we recommend the Jaccard distance as a vertex-to-vertex distance measure.  For example, we use the NetworkX Jaccard coefficient function in our own work \cite{Networkx}.

\section{Metric Space}
A metric space is a set of points that share a distance function. This function must have the following three properties:
\begin{eqnarray}
	g(x,y) &=& 0 \Leftrightarrow x=y \\ 
	g(x,y) &=& g(y,x) \\
	g(x,z) & \le & g(x,y) + g(y,z) \, .
\end{eqnarray}
In the case of the Jaccard distance, the first two properties are immediately apparent. They are direct consequences of the definitions of set operations. The third property, the triangle inequality, was shown to hold by Levandowsky and Winter \cite{Jacc71}.

\section{Conclusion}
We have shown that Jaccard and Otsuka-Ochiai distances, when averaged over clusters provide very accurate estimates of (inverse) intra-cluster density. The Pearson correlation coefficients between these distances and intra-cluster density is almost inversely perfect ($\approx -1$). A visual inspection of the relationship between these distances and intra-cluster density also reveals a perfectly linear inverse relationship.

\bibliography{myBib}

\end{document}